# Hematopoietic cancers and Nup98 fusions: determining common mechanisms of malignancy

**Juliana S. Capitanio and Richard W. Wozniak**
Department of Cell Biology, Faculty of Medicine and Dentistry, University of Alberta

**Abstract:** Chromosomal aberrations are very frequent in leukemias and several recurring mutations capable of malignant transformation have been described. These mutations usually occur in hematopoietic stem cells (HSC), transforming them into leukemia stem cells. NUP98 gene translocations are an example of such chromosomal aberrations; these translocations produce a fusion protein containing the N-terminal portion of Nup98 and the C-terminal of a fusion partner. Over 75% of Nup98 fusions can interact with chromatin, and lead to changes in gene expression. Therefore, I hypothesize that nup98 fusions act as rogue transcriptional regulators in the cell.
Collecting previously published gene expression data (microarray) from HSCs expressing Nup98 fusions, we can generate data to corroborate this hypothesis. Several different fusions affect the expression of similar genes; these are involved in a few biological processes in the cell: embryonic development, immune system formation and chromatin organization. Deregulated genes also present similar transcription factor binding sites in their regulatory regions. These putative regulatory transcription factors are highly interconnected through protein-protein interactions and transcriptional regulation among themselves, and they have important roles in cell cycle regulation, embryonic development, hematopoiesis, apoptosis and chromatin modification.

## Background:

Leukemia is a type of hematologic malignancy that originates in the bone marrow and leads to accumulation of immature hematopoietic cells with clonal origin. These leukemic cells can out-compete normal blood cells, replacing them in the bone marrow and spreading to extramedullary sites, and thus interfering with the normal function of the hematological tissue. The Canadian Cancer Society estimates 5000 new cases of leukemia for 2011, with approximately 2500 deaths.(1)
Leukemia is a heterogeneous disease that can be subdivided according to the cell lineage affected (myeloid or lymphoid) and the degree of differentiation of leukemic cells. More recent classifications by the World Health Organization also incorporate pathologic and genetic markers, achieving more biological significance.(2)

### Chromosomal aberrations and leukemogenesis: delineating causes and effects.

Our knowledge of leukemogenesis is greatly influenced by the discovery of recurring chromosomal aberrations and/or gene mutations capable of malignant transformation of cells.(3) The target cell for these mutations is not always known, but increasing evidence indicates that leukemias originate in hematopoietic stem cells (HSCs) that are transformed into leukemia stem cells (LSC) by these chromosomal aberrations. LSCs and HSCs share two important characteristics, self-renewal and differentiation of new hematopoietic tissue. The clonogenic nature of leukemias is similar to that observed in normal hematopoiesis, and only a small specific subset of leukemic cells, LSCs, are capable of indefinite proliferation, as seen only for HSCs.(4)
If we consider HSCs as the cells of origin for most leukemias, we can better comprehend the high incidence of chromosomal aberrations present in this type of cancer. In a normal organism HSCs are usually quiescent and cycle very slowly; this is a protective mechanism to minimize DNA replication errors and the generation of toxic metabolic subproducts. However, quiescence of HSCs also presents a disadvantage. Once DNA damage does occur, specifically DNA double strand breaks (DSB), these cells must repair the lesions through the non-homologous end joining (NHEJ) pathway. NHEJ is an error-prone, mutagenic pathway that often causes chromosomal aberrations.(5) Given the self-renewal property of HSCs, mutations can be transmitted and accumulated, giving rise to LSCs with a high frequency of chromosomal aberrations; in fact, most recurring translocations in leukemia display a NHEJ repair signature.(6)
Most chromosomal translocations in leukemias occur in chromosomal regions where the DNA is more susceptible to double strand breaks. These susceptibility regions can be characterized by several chromatin structural elements, including topo II DNA cleavage sites, DNase I hypersensitive sites, scaffold/matrix attachment regions (S/MAR) and retrotransposon regions (LINE and SINE).(6) Chromosomal position within the nucleus also influences the frequency of translocations. It is likely that loci localized in close proximity inside the nucleus will more frequently translocate in cases of DSBs.(7) Finally, for these translocations to lead to transformation, they must generate fusion proteins that promote an advantage to the cell.
Thousands of chromosomal translocations have been detected in leukemia cells.(8) Most of these fall within two categories: type I mutations promote increased proliferation or survival, and type II mutations impair differentiation or enhance self-renewal. At least one mutation of each type seems to be necessary for leukemogenesis.(9) Leukemic chromosomal translocations generate chimeric fusion proteins, and many of these fusion proteins have similar characteristics: they localize to the nucleus, affect transcriptional regulation, contain a DNA binding domain and cause epigenetic modifications.(10)
In fact, epigenetic changes are a common occurrence in most acute leukemias.(3) If we consider that during normal hematopoiesis a complex program of epigenetic modifications takes place, it becomes clearer how altering epigenetic modifications can affect cell differentiation and self-renewal leading to leukemogenesis.(11) HOX genes are a prime example of epigenetic regulation. The pattern of HOX expres-sion in cells is epigenetically regulated and inherited; each cell in the hematopoietic differentiation continuum displays a specific pattern of HOX genes expressed. More primitive hematopoietic cells have higher levels of HOX expression, as cells differentiate and lose their proliferative capabilities, HOX expression decreases until it becomes absent in completely differentiated cells. Overexpression of numerous HOX genes can induce leukemogenesis, in several cases changes in their expression levels are a result of histone modifications in the 5' HOXA gene.(12)

## Nucleoporin genes and cancer.

To date, the literature has revealed five nucleoporins involved in carcinogenesis, Tpr, Nup88, Nup98, Nup214(13) and Nup358.(14) Nup88 expression is up-regulated in several cancers, specially carcinomas. Increase in its protein levels is thought to deregulate NF-kB nuclear transport maintaining it constantly activated.(15) The remaining 4 nucleoporins are involved in carcinogenic gene fusions.

Tpr gene fusions with Met and NTrk1 have been described in gastric cancers and papillary thyroid carcinomas, respectively. In both cases the N-terminal coiled-coil domain of Tpr is juxtaposed with the tyrosine kinase domain of the fusion partner. This leads to dimerization independent of ligand, and to constitutive activation of kinase activity, causing deregulated signaling that leads to carcinogenesis.(13) A similar carcinogenic mechanism is seen in Nup358 fusions. In inflammatory myofibroblastic tumors, the N-terminal leucine zipper of Nup358 is fused to the tyrosine kinase domain of ALK, also leading to deregulated kinase activation.(16)

Gene fusion of NUP214 and NUP98 play a role in leukemogenesis. NUP214 gene fusions with ABL also promote constitutive kinase activation in T-ALL. The kinase domain of ABL is fused to the N-terminal coiled-coil motif of Nup214. Interaction between this motif and nup88 allows localization of the fusion to the nuclear pore complex (NPC), bringing the kinase domains to sufficient proximity for constitutive activation.(17) In very rare cases Nup214 also translocates with SET and DEK in T-ALL and AML, respectively. In these cases almost the full length SET or DEK protein is fused to the C-terminal FG repeat of Nup214.(13) SET-NUP214 can interact with HOXA gene promoters leading to their expression. HOXA expression only occurs in the earliest T-cell precursor, so the fusion blocks T cell differentiation.(18) DEK-NUP214 fusions seem to increase overall protein translation specifically in myeloid cells, possibly facilitating carcinogenesis.(19) SET and DEK are two histone interacting proteins that perform opposing roles in the regulation of access to chromatin. SET promotes and DEK restricts accessibility to chromatin by the transcriptional machinery.(20) It is possible that NUP214 translocation with these genes affects the balance between the two chromatin modifiers.

The NUP98 gene is fused to a wide range of partner genes, resulting in several hematopoietic disorders, especially acute myeloid leukemia (AML).

## NUP98 translocations in leukemias.

All NUP98 translocations described thus far generate a chimeric fusion protein that retains the N-terminal portion of Nup98 and the C-terminal of the fusion partner. Most chromosomal breaks take place between exons 11 and 13 of the NUP98 gene.(21) Interestingly, an enrichment of DNAse I hypersensitivity sites and a strong prediction of S/MAR are present in this region, corroborating its increased susceptibility to translocations, as seen in figure 1 below.

A

B

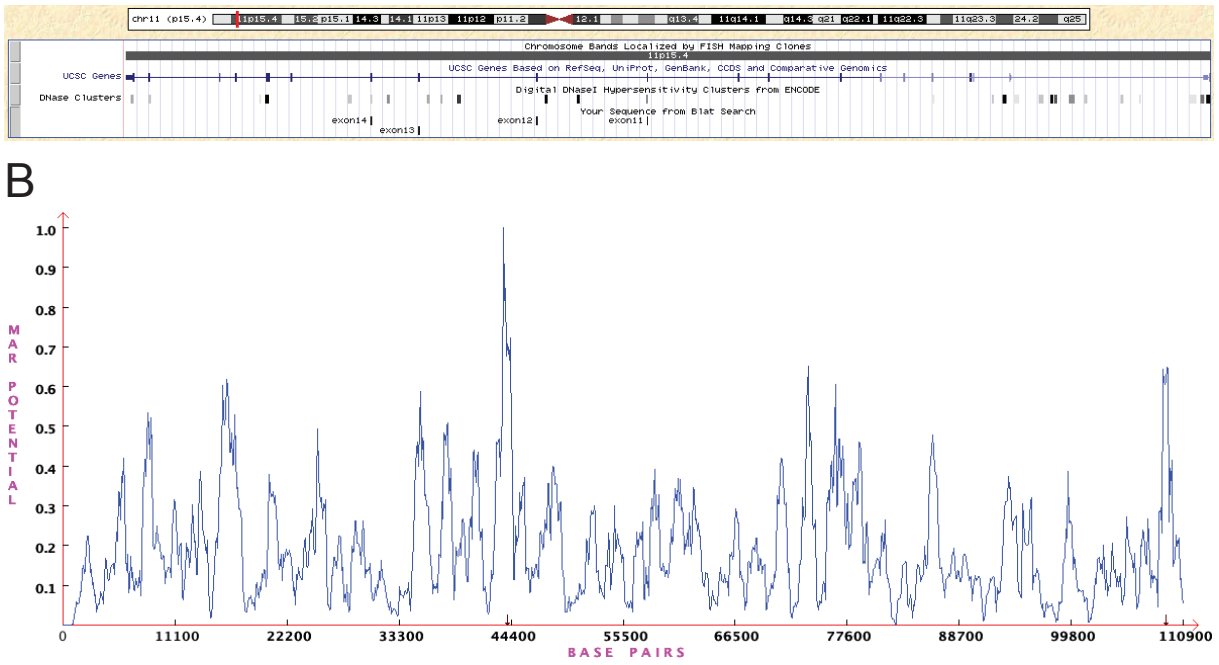

**Figure 1: NUP98 gene chromatin structure.** A – UCSC genome browser(52) tracks displaying NUP98 gene structure (most commonly translocated region, exons 11 to 14, marked) and DNAse I hypersensitivity sites. B – S/MAR enrichment on NUP98 gene, defined by MARFinder.(53)

The N-terminal portion of Nup98 (conserved in fusions) contains FG/GLFG repeats flanking a coiled-coil Rae1 interaction site (Fig. 2A). Even though a third of all nucleoporins contain FG repeats, Nup98 is the only GLFG repeat containing nucleoporin in humans.(22) At nuclear pore complexes (NPC), Nup98 interacts with transport molecules, mediating traffic through the NPC. The N-terminal of Nup98 interacts with XPO1, facilitating export of specific proteins from the nucleus,(23) and with TAP and Rae1, promoting mRNA export to the cytoplasm.(24)

Nup98 can also be found away from the NPC, dispersed through the nucleoplasm and in intranuclear structures called GLFG bodies (Fig. 2B).(25) In embryonic Drosophila cells, the intranuclear pool of Nup98 interacts with transcriptionally active genes and changes to the level of Nup98 present can modulate their expression, especially in developmental genes.(26) In human cells, the N-terminal GLFG repeats of nup98 have been shown to interact with histone acetyl transferases and histone deacetylases.(27, 28)

At least 27 different genes have been found translocated with NUP98 in leukemic patients (Fig. 2C). Most of these gene fusions lead to myeloid malignancies (AML, CML, MDS); however, six fusions have so far been identified in T-ALL patients. Nup98 fusions are rare (approximately 2% AML cases) however, they usually indicate a poor prognosis. Over half of Nup98 fusions are detected in patients under 20 years of age and only 25% occur in patients with therapy related malignancies.(21) The karyotype of patients with Nup98 fusions is usually simple, with no more than 3 chromosomal aberrations,(29) indicating a strong transformation potential for the fusions and arguing against an increase in genetic instability.

Nup98 fusion partners can be divided into homeodomain (HD) and non-HD containing proteins. Fusions with HD containing proteins always maintain the N-terminal GLFG domain of Nup98 fused in frame to the C-terminal HD of the partner gene.(30) NUP98-HOXA9 was the first fusion detected in an AML patient(31, 32) and it is currently the best-characterized Nup98 translocation. All non-HD containing partner genes encode putative coiled-coil motifs,(33) a domain usually involved in mediating protein-protein interactions. Chromatin recognition domains, such as plant homeodomain zinc fingers (PHD), are also recurrent in non-HD Nup98 fusion partners.(34)

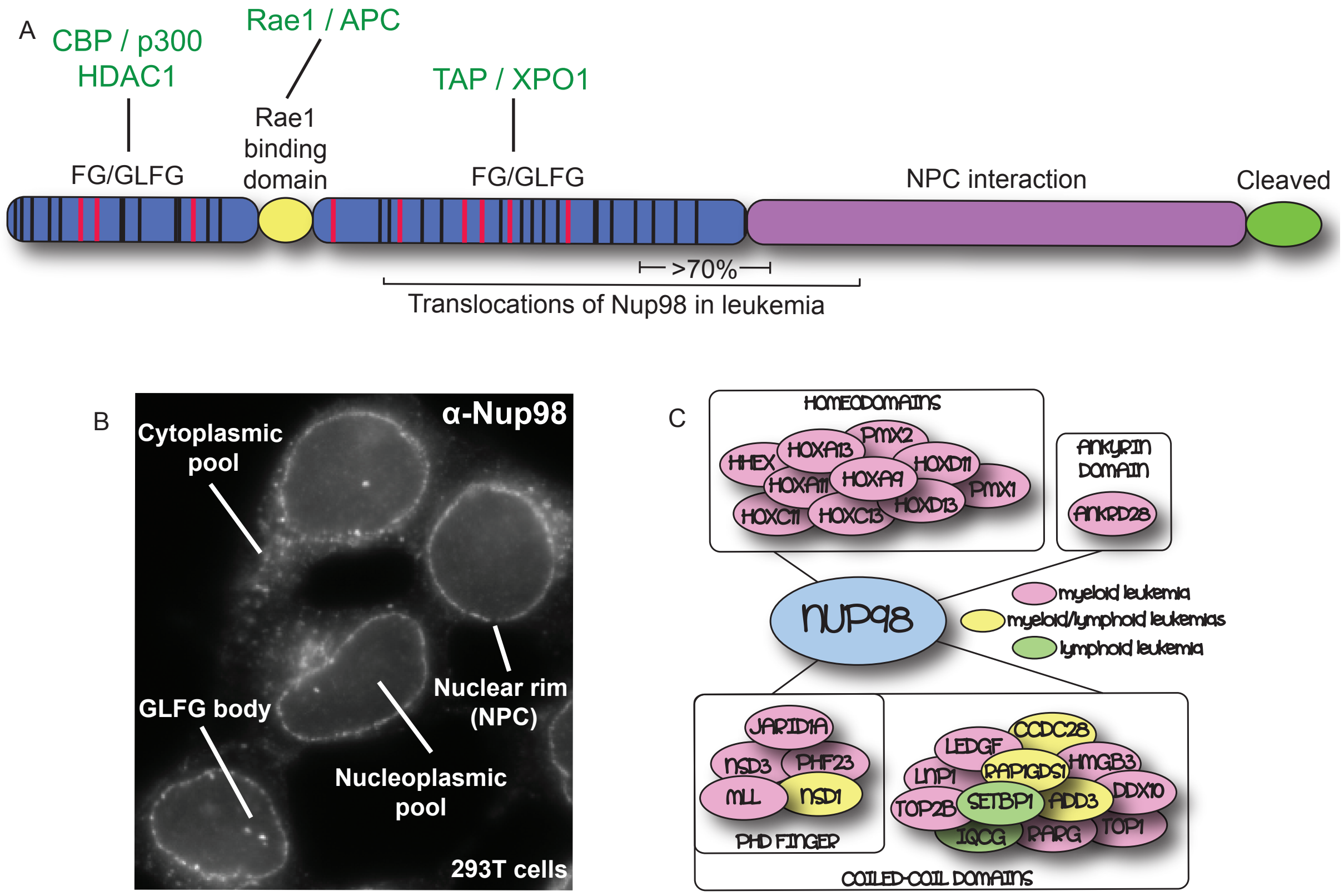

**Figure 2: Nup98 and Nup98 fusions.** A – Nup98 protein domains and known interactions, B – Nup98 localization in 293T cells visualized by immunofluorescence microscopy and C – Diagram of know Nup98 gene fusions and their characteristics.

So far, few common denominators have been identified when it come to the mechanism by which Nup98 fusions may lead to leukemogenesis. Characterizing their effects in altered gene expression indicates a few common targets: fusions of Nup98 with NSD1, KDM5A, PHF23, HOXA9, HOXD13, PRRX1, HHEX and DDX10 seem to increase the expression of HOXA cluster genes; NSD1, HOXA9 and DDX10 fusions also up-regulate the Hox co-factor Meis1; and HOXA9 or HOXD13 translocations increase the expression of interferon responsive genes.(21) A putative mechanism for how these fusions may alter gene expression has only been described for PHD domain containing translocations (NSD1, KDM5A and PHF23). These fusions seem to bind HOXA gene promoters (through the PHD finger) and recruit histone acetylases CBP/p300 (via GLFG domain) that modify chromatin into a transcriptionally active state.(34, 35) Nup98 fusions with HD containing proteins are assumed to act directly as transcription factors; they can in some cases collaborate with Meis1, a Hox co-factor that increases specificity and binding to target DNAs.(36)

## Preliminary analysis:

### Commonalities among NUP98 translocation partner genes.

Performing bioinformatics analysis of Nup98 fusion partner genes, a few recurrent themes are uncovered (Tab. I). Investigating the interaction profile of these partner proteins we notice that chromatin and/or DNA binding is a characteristic of 75% of them. The partner genes usually display direct DNA interaction or recognition of histone post-translational modifications, with 50% of all fusion partners working as transcriptional regulators. Separating Nup98 fusions leading to myeloid malignancies from those causing T-ALL, an even clearer picture appears: over 95% of myeloid related fusions can interact with DNA/chromatin, with over 2/3 of them acting as transcriptional regulators. T-ALL related fusions show no transcriptional regulators among partner genes and only one of them can interact with DNA/chromatin, likely pointing towards distinct molecular mechanisms for myeloid and lymphoid causing Nup98 fusions.

Evaluating biological processes affected by Nup98 fusion partner genes, we determine that almost 60% of them participate in embryonic regionalization and development, with over half being involved in transcription. Other biological processes over-represented among the fusion partners are regulation of cell proliferation, cell differentiation and chromatin modification (Tab. I). Interestingly, in Drosophila cells, Nup98 itself seems to regulate the transcription of developmental and cell cycle genes.(26)

The data above indicate that Nup98 fusions might function as rogue transcriptional regulators, especially in myeloid malignancies. It is possible that these fusions can affect gene expression acting directly as transcription factors (TF), altering histone modifications or deregulating other TFs.

## Hypothesis:

The goal of this project is to define common mechanisms by which these different Nup98 fusions lead to malignancy. Based on the background and on the preliminary analysis provided above we hypothesize that Nup98 fusions, especially those leading to AML, might function as rogue tran-scriptional regulators, and that their deregulated target genes might impair cell differentiation and increase self-renewal, setting the stage for malignant transformation and acute myeloid leukemia.

**Table I: Characteristics of Nup98 fusion partner genes.**

| | | Nup98 fusion genes | | | | | | | | | | | | | | | | | | | | | | | | | | | | | |
|---|---|---|---|---|---|---|---|---|---|---|---|---|---|---|---|---|---|---|---|---|---|---|---|---|---|---|---|---|---|---|---|
| | | ADD3 | CCDC28A | DDX10 | HHEX | HMGB3 | HOXA11 | HOXA13 | HOXA9 | HOXC11 | HOXC13 | HOXD11 | HOXD13 | IQCG | KDM5A | LNP1 | MLL | NSD1 | PHF23 | PRRX1 | PRRX2 | PSIP1 | RAP1GDS1 | RARG | SETBP1 | TOP1 | TOP2B | WHSC1L1 | Total | AML | ALL |
| Interaction | DNA/Chromatin binding | | | | | | | | | | | | | | | | | | | | | | | | | | | | 77.78% | 95.24% | 16.67% |
| | Transcription regulator | | | | | | | | | | | | | | | | | | | | | | | | | | | | 51.85% | 66.67% | 0.00% |
| Protein domains | Homeobox | | | | | | | | | | | | | | | | | | | | | | | | | | | | 35.71% | 40.00% | 0.00% |
| | Coiled-coil | | | | | | | | | | | | | | | | | | | | | | | | | | | | 62.96% | 52.38% | 100.00% |
| | Zinc finger PHD | | | | | | | | | | | | | | | | | | | | | | | | | | | | 17.86% | 20.00% | 0.00% |
| Biological Process | Embryonic regionalization | | | | | | | | | | | | | | | | | | | | | | | | | | | | 59.26% | 57.14% | 66.67% |
| | Cell differentiation | | | | | | | | | | | | | | | | | | | | | | | | | | | | 18.52% | 23.81% | 0.00% |
| | Regulation of cell proliferation | | | | | | | | | | | | | | | | | | | | | | | | | | | | 22.22% | 19.05% | 33.33% |
| | Transcription | | | | | | | | | | | | | | | | | | | | | | | | | | | | 51.85% | 47.62% | 66.67% |
| | Chromatin modification | | | | | | | | | | | | | | | | | | | | | | | | | | | | 14.81% | 9.52% | 33.33% |
| Lineage | Myeloid | | | | | | | | | | | | | | | | | | | | | | | | | | | | 77.78% | | |
| | Lymphoid | | | | | | | | | | | | | | | | | | | | | | | | | | | | 22.22% | | |

In this project, we propose to study the changes in gene expression caused by Nup98 translocations in bone marrow cells. Using data integration of previously published microarray experiments we will compare the effects of different Nup98 fusion proteins in the gene expression profile of bone marrow cells, leading to the discovery of specific pathways responsible for the disease phenotype. These relevant pathways can indicate key drug targets for this malignancy, and drug responses can be modeled in the existing networks, aiding in the development of new therapies for this disease.

## Materials and Methods:

In order to further explore the possible role of Nup98 fusions as rogue transcriptional regulators, I collected microarray experiments of bone marrow cells transformed with different fusions for analysis. In order for this multi-experiment analysis to present biological significance, I only used results from experiments performed in similar conditions, ending up with 4 directly comparable sets: NUP98-HHEX, NUP98-HOXA9, NUP98-HOXA10, NUP98-HOXD13(37, 38).

All fusions were transduced into adult mice bone marrow cells using the retroviral vector MSCV-IRES-GFP (the empty vector was used as control), cells were FACS sorted before mRNA purification, target preparation and hybridization to Affymetrix Mouse Genome 430A Arrays. The raw (.CEL) file of each experiment was RMA (robust multichip average) normalized(39) and cross-study normalization was achieved using ComBat,(40) an empirical bayes method. The ANOVA statistical test ($p<0.05$) was used in AltAnalyze(39) to identify genes differentially expressed in control vs. Nup98 fusion samples, producing a list of genes whose expression was similarly affected by all Nup98 fusion proteins.

Functional annotation analysis of all gene lists was performed through the Database for Annotation, Visualization and Integrated Discovery (DAVID)(41) v6.7. Distant regulatory elements of co-regulated genes, such as transcription factor binding sites and CpG islands, were identified using DiRE.(42) Finally, functional protein association networks were inferred using STRING(43) version 9.0 and were imported into Cytoscape(44) for formatting.

## Results and Discussion:

### The expression of several genes is similarly altered in the presence of different Nup98 fusions.

Analyzing the genes whose expression was affected by all Nup98 fusions similarly, we see enrichment of a few biological processes: embryonic development, immune system formation and chromatin organization (Fig. 3A and Tab. S1). Evaluating only those genes whose expression was increased in the presence of all NUP98 translocations, we see enrichment for regulation of transcription, cell proliferation and immune system development (Fig. 3B). On the other hand, genes with decreased expression in the presence of Nup98 fusions are overrepresented for embryonic development, RNA processing and chromatin modification (Fig. 3C). These changes in the expression profile of genes involved in chromatin organization and modification can be correlated to the know epigenetic deregulation occurring in leukemic cells, especially as has been described for cells containing NUP98 translocations.(21) An increase in the expression of cell proliferation genes can also explain the expansion in the number of these LSCs. They abandon quiescence, as seen in HSCs, and actively proliferate in a deregulated manner, contributing to malignancy.

### Genes with altered expression in the presence of NUP98 fusions are regulated by similar transcription factors.

Mapping the regulatory regions and transcription factor binding sites (TFBS) present in the deregulated genes, we can see that up and down-regulated genes present several regulatory regions in common, and 38 transcription factor binding site are enriched in both sets of genes (Tab. 2).

These transcription factors play relevant roles in cell cycle regulation, embryonic development, hematopoiesis, apoptosis and chromatin modifications (Fig. 4). They form a highly interconnected network,(43) indicating protein-protein interactions and transcriptional regulation among themselves (Fig. 5). Networks of TFs regulating genes with increased expression or genes with decreased expression are less interconnected than that of TFs that are present in both; however,

**Table 2: Transcription factors with TFBS enriched in genes with deregulated expression in the presence of nup98 fusions.**

| Common TFs | Symbol | Gene ID | Occurence | | Importance | |
|---|---|---|---|---|---|---|
| | | | up reg genes | down reg genes | up reg genes | down reg genes |
| CEBP | cebpa | 12606 | 8.82% | 4.66% | 0.01544 | 0.0459 |
| | cebpb | 12608 | | | | |
| CMAF | maf | 17132 | 4.41% | 1.43% | 0.0024 | 0.00896 |
| CRX | crx | 12951 | 2.94% | 1.08% | 0.00037 | 0.00329 |
| DEC | Bhlhe40 | 20893 | 7.35% | 2.87% | 0.07077 | 0.00621 |
| E2A | tcf3 | 21423 | 8.82% | 7.17% | 0.02206 | 0.00341 |
| E2F1DP1 | e2f1 | 13555 | 5.88% | 2.15% | 0.06655 | 0.00324 |
| | tfdp1 | 21781 | | | | |
| E2F1DP1RB | e2f1 | 13555 | 5.88% | 2.87% | 0.03318 | 0.01098 |
| | tfdp1 | 21781 | | | | |
| | rb1 | 19645 | | | | |
| E2F4DP1 | e2f4 | 104394 | 5.88% | 2.87% | 0.00147 | 0.00229 |
| | tfdp1 | 21781 | | | | |
| FOXJ2 | foxj2 | 60611 | 7.35% | 1.79% | 0.13281 | 0.01915 |
| FOXO4 | foxo4 | 54601 | 7.35% | 1.08% | 0.00184 | 0.00184 |
| GC | CpG islands | | 16.18% | 19.35% | 0.03808 | 0.13548 |
| GCM | gcm2 | 107889 | 7.35% | 4.30% | 0.06147 | 0.03602 |
| | gcm1 | 14531 | | | | |
| GRE | glucocorticoid res. elem. | | 1.47% | 1.08% | 0.0136 | 0.00197 |
| HFH1 | foxq1 | 15220 | 2.94% | 0.36% | 0.01454 | 0.00205 |
| HIF1 | hif1a | 15251 | 11.76% | 9.68% | 0.02061 | 0.04876 |
| HNF1 | hnf1a | 21405 | 8.82% | 2.51% | 0.03529 | 0.01356 |
| | hnf1b | 21410 | | | | |
| IK1 | ikzf1 | 22778 | 4.41% | 2.15% | 0.02509 | 0.00403 |
| LXR | Nr1h2 | 22260 | 7.35% | 2.15% | 0.07146 | 0.01984 |
| | nr1h3 | 22259 | | | | |
| MYB | myb | 17863 | 7.35% | 2.51% | 0.02941 | 0.00044 |
| MYC | myc | 17869 | 1.47% | 2.87% | 0.01418 | 0.02151 |
| MYOGENIN | myog | 17928 | 7.35% | 6.09% | 0.05285 | 0.00515 |
| NERF | elf2 | 69257 | 8.82% | 5.02% | 0.09816 | 0.01944 |
| NFKAPPAB50 | nfkb1 | 18033 | 7.35% | 2.51% | 0.16176 | 0.00659 |
| NFKB | rela | 19697 | 8.82% | 6.81% | 0.02105 | 0.06719 |
| NFY | nfia | 18027 | 13.24% | 4.66% | 0.03174 | 0.00453 |
| | nfib | 18028 | | | | |
| | nfic | 18029 | | | | |
| | nfix | 18032 | | | | |
| P53 | trp53 | 22059 | 8.82% | 1.43% | 0.14366 | 0.00409 |
| PAX | pax1 | 18503 | 5.88% | 2.51% | 0.04559 | 0.01687 |
| PAX9 | pax9 | 18511 | 4.41% | 2.87% | 0.01665 | 0.01093 |
| RSRFC4 | mef2a | 17258 | 2.94% | 0.72% | 0.00126 | 0.00905 |
| SREBP | srebf1 | 20787 | 5.88% | 2.87% | 0.03676 | 0.0145 |
| | srebf2 | 20788 | | | | |
| STAT | stat1 | 20846 | 7.35% | 2.51% | 0.0432 | 0.01443 |
| TAL1 | tal1 | 21349 | 5.88% | 2.15% | 0.06324 | 0.00225 |
| TCF4 | tcf4 | 21413 | 2.94% | 2.15% | 0.04403 | 0.01667 |
| TEL2 | telo2 | 71718 | 1.47% | 2.15% | 0.00729 | 0.01726 |
| TST1 | pou3f1 | 18991 | 1.47% | 1.08% | 0.01556 | 0.0121 |
| USF2 | usf2 | 22282 | 1.47% | 2.87% | 0.00023 | 0.02222 |
| WT1 | wt1 | 22431 | 14.71% | 6.45% | 0.03808 | 0.03055 |
| ZF5 | zfp161 | 22666 | 27.94% | 23.30% | 0.08507 | 0.13499 |
| OCT4 | pou5f1 | 18999 | 2.94% | 1.08% | 0.01838 | 0.00753 |

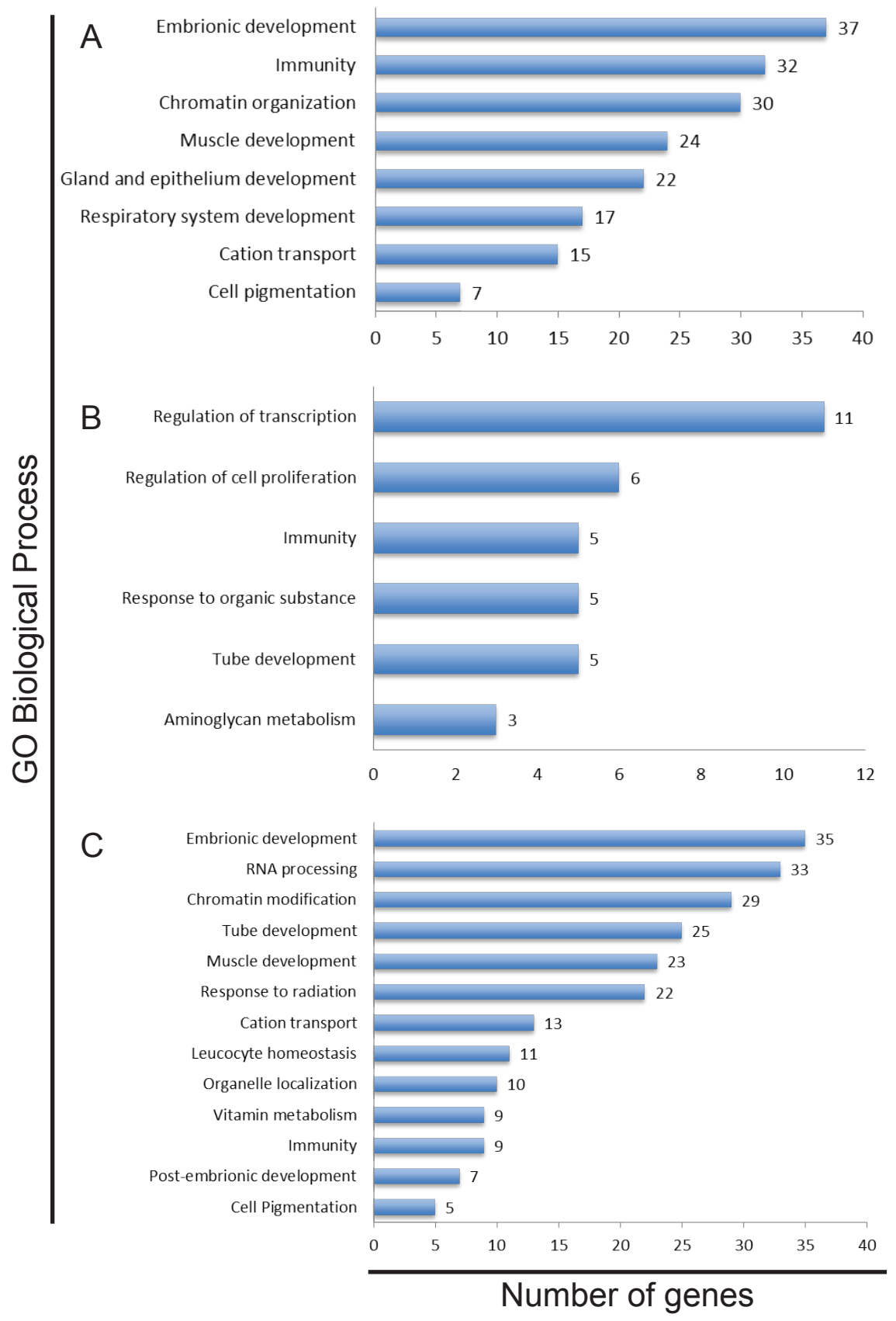

**Figure 3: GO annotation biological process enrichment** A) all genes differentially expressed; B) up-regulated genes; C) down-regulated genes in the presence of Nup98 fusion proteins.

each network contains at least one module of a few highly interconnected TFs (Fig. S1 and S2).

### Some of these transcription factors are known to also be deregulated in other types of leukemia.

More in-depth analysis of the role of a few of the TFs putatively regulating genes with altered expression in cells bearing NUP98 translocations can uncover possible novel mechanisms by which these fusions may lead to leukemogenesis. Wilms Tumor 1 (WT1) is one of the most enriched TFBS in up and down regulated genes. WT1 is a TF with expression restricted to hematopoietic progenitor cells in the bone marrow with a role in their self-renewal. Mutations in WT1 indicate a worse prognosis in acute leukemias and can be found in approximately 10% of AML cases.(45) This transcription factor can work as both a tumor suppressor and an oncogene, and it can enhance or repress transcription of its target genes (such as MYC and BCL-2) depending on cellular conditions.(46) WT1 and N-terminal Nup98 (present in fusions) both interact with CBP, providing an interesting putative mechanism on how Nup98 fusions might be affecting this TF and its targets without affecting its expression level.

As another example, CEBPA appears as an important regulatory TF in genes deregulated by Nup98 fusions. Interestingly its own expression is reduced nearly 2 fold in Nup98 fusion expressing cells. Decreased CEBPA expression in BM cells decreases differentiation and increases proliferation of myeloid progenitors leading to leukemia. Another leukemic fusion, AML1-ETO, has also been shown to down-regulate CEBPA expression, and CEBPA mutations that abrogate its function or generate dominant negatives have also been described as leukemogenic.(47)

TFBS for MYC are also overrepresented in this deregulated gene set, albeit to a lesser extent. MYC expression increases 40% in the presence of Nup98 fusions, similarly to what is observed with several other leukemic chromosomal aberrations (AML1-ETO, PML-RARA, PLZF-RARA, FLT3-ITD) shown to induce c-myc activation. Overexpression of c-myc alone in BM cells can quickly induce fatal AML.(48)

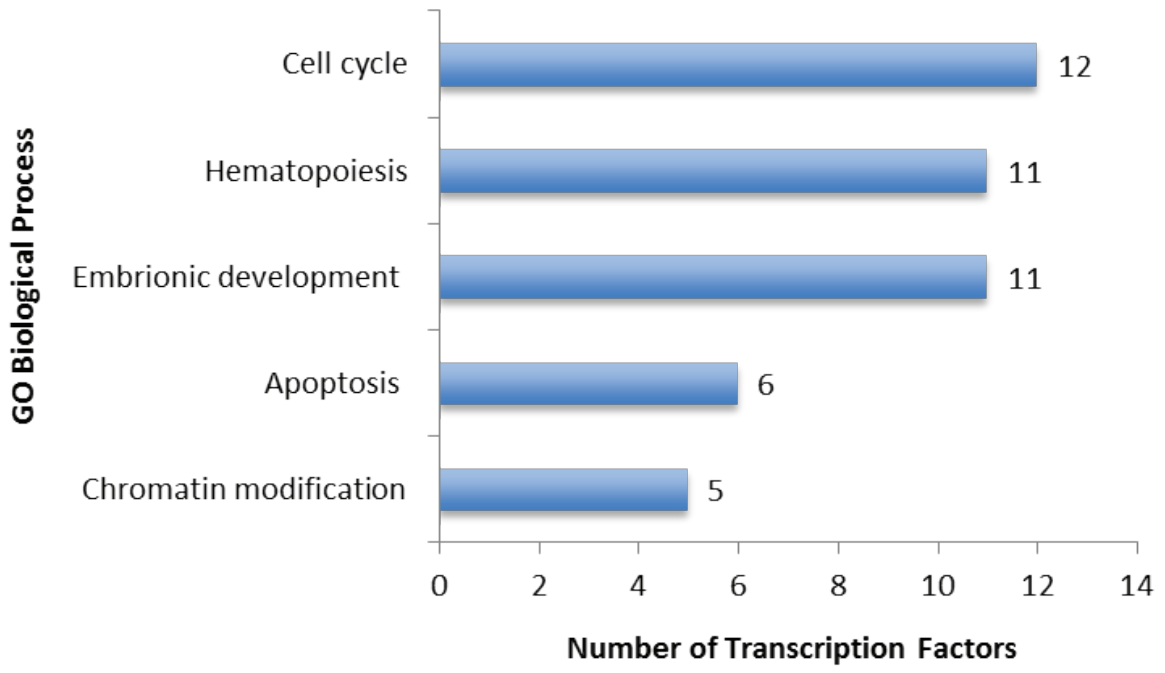

**Figure 4: GO annotation biological process enrichment for all TFs** with TFBS enriched in genes differentially expressed in the presence of Nup98 fusion proteins.

### The identified transcription factor network is highly similar to the network of transcription factors regulating growth arrest and differentiation in human myeloid cells.

Given the above results, we can hypothesize that these NUP98 translocations are deregulating transcription factors that control differentiation and self-renewal in primitive hematopoietic cells. The network of transcription factors regulating growth arrest and differentiation in a human myeloid cell line has already been described.(49) Superposing the microarray results obtained above into this previously published network (Fig. 6), we notice that over 80% of the TFs thought to regulate growth arrest and differentiation have decreased expression in cells containing Nup98 fusions. Additionally, 70% of the TFs represented in this network have enriched TFBS in genes with deregulated expression upon NUP98 translocations. This reinforces the idea that Nup98 fusions can deregulate key TFs in myeloid cells, leading to a cascade of changes in their gene expression profile that ultimately disrupts differentiation and proliferation, promoting leukemogenesis.

### Enrichment of CpG islands in the promoter region of genes with deregulated expression further indicates epigenetic deregulation in the presence of Nup98 fusions.

A final interesting observation is the enrichment of CpG islands in the promoters of genes with deregulated expression in Nup98 fusions, indicating an important role for epigenetic changes in their altered expression. Epigenetic regulation of gene expression is a hallmark of hematopoiesis, and several leukemic translocations have been shown to alter transcription by altering epigenetic markers in their target genes.(3) HOX genes are an example of this epigenetic regulation and they are highly enriched for CpG islands in humans.(50) Several Nup98 fusions promote HOX genes up-regulation,(21) some of them have been shown to alter histone post transla-tional modifications in the HOX locus,(34, 35) and it is possible that Nup98 fusions may also lead to epigenetic deregulation affecting DNA methylation or histone modifications in the CpG islands identified in deregulated genes.

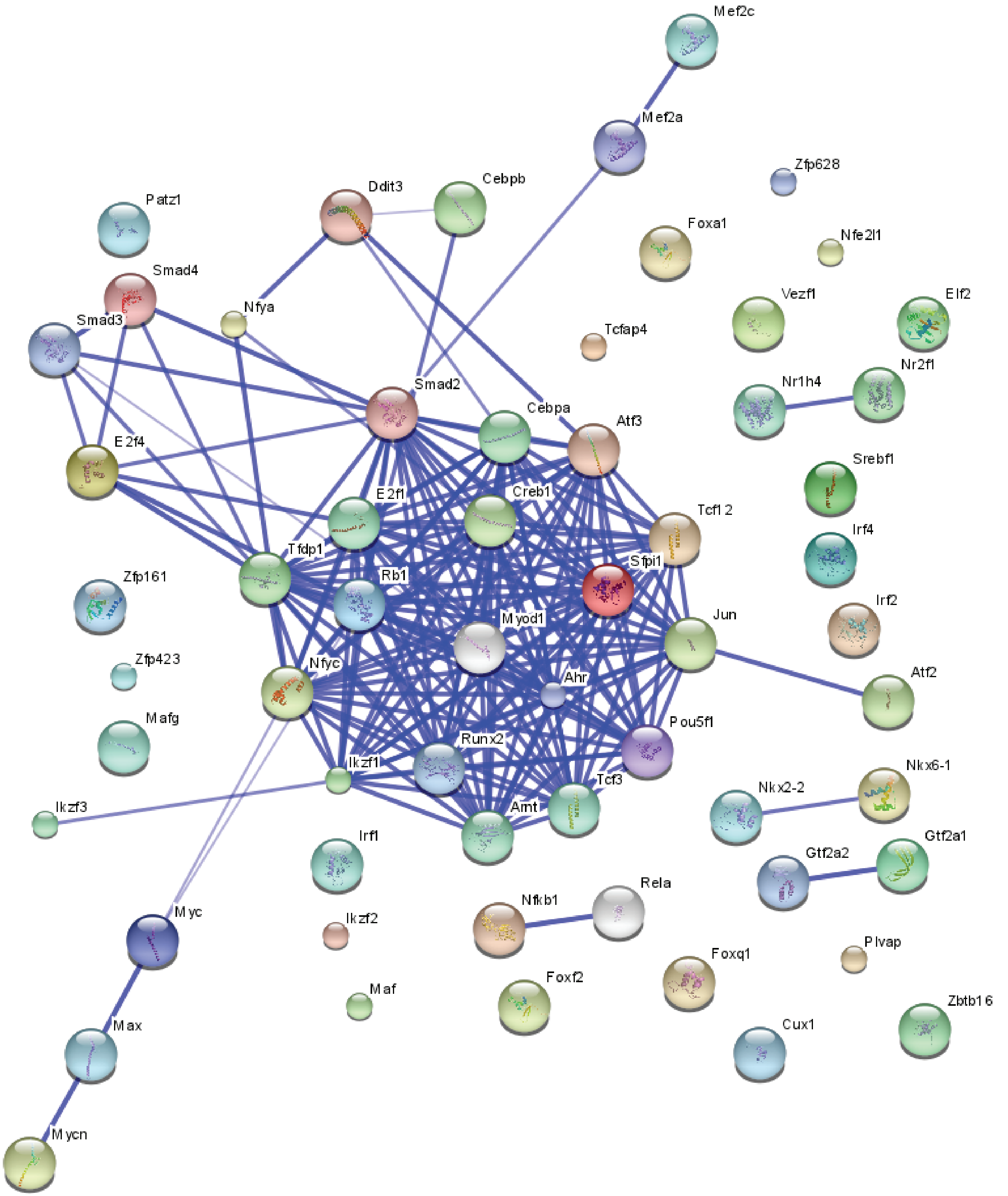

**Figure 5: Protein-protein interaction network of TFs** with TFBS enriched in genes differentially expressed in Nup98 translocations.

As well as DNA methylation, CpG islands in gene promoters can be silenced by polycomb group proteins.(51) In mouse hematopoietic progenitors NUP98-KDM5A fusions bind the promoters of HOXA6-A10, this stops polycomb complex proteins from silencing them, maintains H3K4me3 and acetylated histones in their promoter, and leads to increased expression of these genes.(34) Similarly, NUP98-NSD1 binds the promoters of HOXA7 and HOXA9, leading to histone acetylation (p300 recruitment by Nup98's N-terminal) and H3K36me3 (NSD1 C-terminal function) that increase gene expression.(35) It's possible that other Nup98 fusion proteins may have similar mechanisms of epigenetic deregulation.

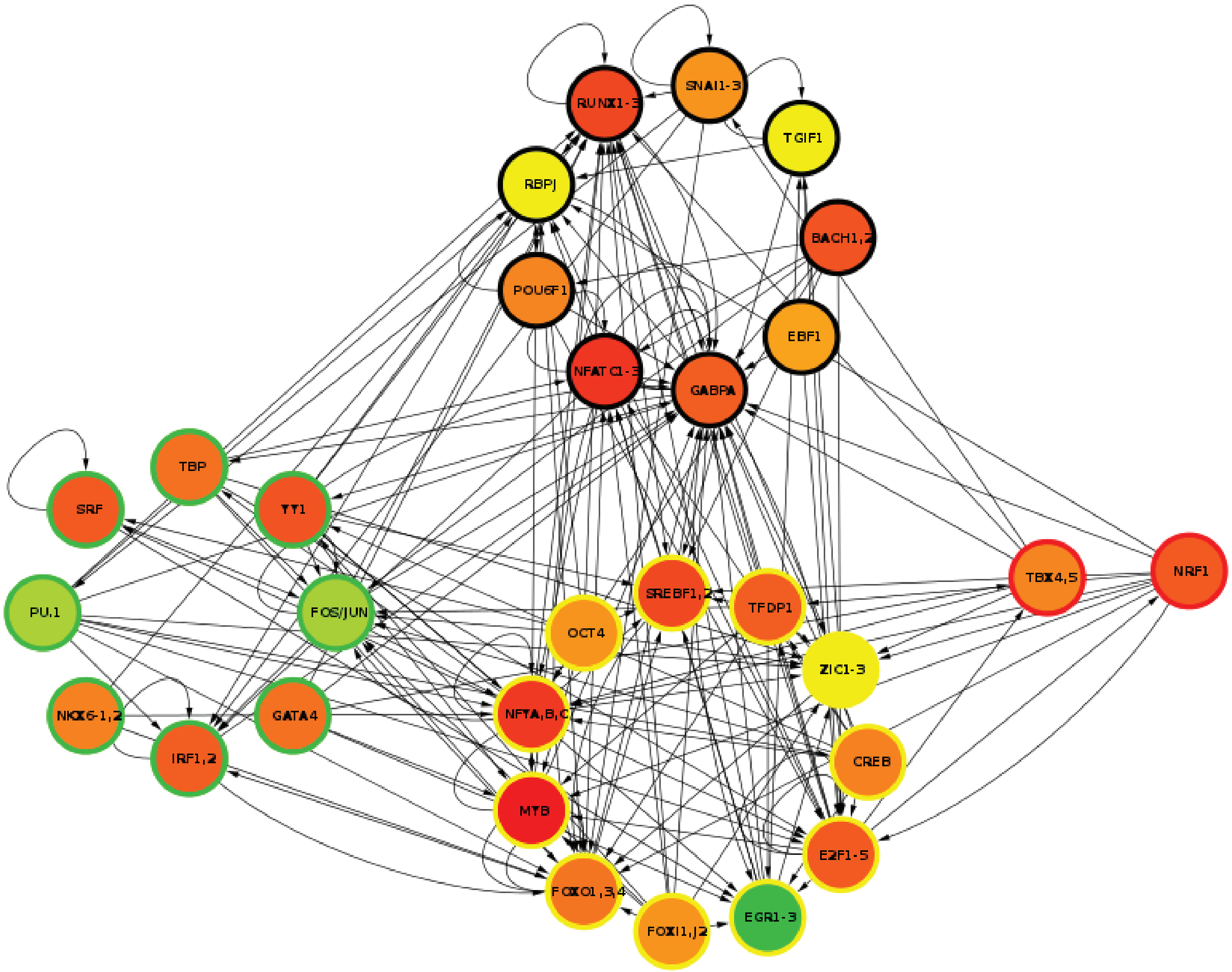

**Figure 6: Transcriptional network of growth arrest and differentiation in a human myeloid leukemia cell line.**(49) Superposition of results from Nup98 fusion microarrays into the previously published network of growth arrest and differentiation. Node color indicates expression level of TF (red – decreased, green – increased) and node border indicates enrichment of TFBS in groups of genes with increased (green) or decreased (red) expression (yellow border - TFBS present in up and down regulated genes, black – TFBS absent in deregulated genes).

## Concluding remarks:

The project described here aims at improving our understanding of how NUP98 translocations lead to leukemia. Based on previous literature and the results presented we can see that most Nup98 fusion partners provide a DNA/chromatin interacting interface to these fusion proteins. This leads to the deregulation of sets of genes that increase the proliferation of these cells, as well as to the deregulation of genes that alter epigenetics. The genes with altered expression in the presence of Nup98 fusions are regulated by similar transcription factors, and these form a highly interconnected network. Part of this regulatory transcription factor network is itself down-regulated and very similar to the network of transcription factors regulating growth arrest and differentiation in human myeloid cells. It's likely therefore, that these Nup98 fusions act as rogue transcriptional regulators, affecting mainly cell differentiation and self-renewal.

**Table S1: Genes with altered expression in the presence of Nup98 fusions.**

| Probesets | Symbol | Ensembl_id | Entrez_id | avg-Control [C] | avg-Nup98 Fusions [F] | log_fold-C_vs_F | fold-C_vs_F | rawp-C_vs_F | ANOVA-rawp |
|---|---|---|---|---|---|---|---|---|---|
| 1419770_at | 96298 | ENSMUSG00000089935\| | | 3.038673785 | 2.706183322 | 0.332490463 | 1.25918518 | 0.002200226 | 0.002200226 |
| 1436881_x_at | Afp | ENSMUSG00000054932 | 11576 | 4.435669685 | 4.133741331 | 0.301928354 | 1.2327911 | 0.002303028 | 0.002303028 |
| 1423341_at | Cspg4 | ENSMUSG00000032911\| | 121021 | 4.142224673 | 3.915493832 | 0.226730841 | 1.1701803 | 0.002454198 | 0.002454198 |
| 1427291_at | Sycp1 | ENSMUSG00000027855 | 20957 | 2.864705994 | 2.58908769 | 0.275618304 | 1.21051277 | 0.002480178 | 0.002480178 |
| 1431505_at | LOC629952 /// Ppih | ENSMUSG00000082045\| ENSMUSG00000060288 | 6411\|629 952 | 4.26060466 | 3.910817129 | 0.349787531 | 1.27437293 | 0.002634631 | 0.002634631 |
| 1443473_at | C79562 | | 97158 | 2.999440742 | 2.672116303 | 0.327324439 | 1.25468433 | 0.00302522 | 0.00302522 |
| 1421301_at | Zic2 | ENSMUSG00000061524 | 22772 | 3.200578743 | 2.909484227 | 0.291094516 | 1.2235682 | 0.003413885 | 0.003413885 |
| 1449950_at | 5830415L20Rik | | 68152 | 3.914999249 | 3.580822413 | 0.334176836 | 1.2606579 | 0.003426622 | 0.003426622 |
| 1421741_at | Cyp3a16 | ENSMUSG00000038656 | 13114 | 2.231387797 | 2.044910463 | 0.186477333 | 1.13798168 | 0.003659455 | 0.003659455 |
| 1422607_at | Etv1 | ENSMUSG00000047643\| ENSMUSG00000004151 | 14009 | 3.508771097 | 3.276549487 | 0.23222161 | 1.17464239 | 0.003725371 | 0.003725371 |
| 1428784_at | Gmip | ENSMUSG00000036246 | 78816 | 6.638922179 | 6.095950604 | 0.542971575 | 1.45697041 | 0.003795931 | 0.003795931 |
| 1423410_at | Meig1 | ENSMUSG00000026650 | 104362 | 2.616300359 | 2.358760065 | 0.257540294 | 1.19543882 | 0.003879921 | 0.003879921 |
| 1427603_at | Atf7ip2 | ENSMUSG00000039200\| | 75329 | 2.584209067 | 2.354541658 | 0.229667409 | 1.1725646 | 0.003922985 | 0.003922985 |
| 1455813_at | OTTMUSG0000 0010009 | ENSMUSG00000066031 | 194227 | 2.326065306 | 2.028825087 | 0.297240219 | 1.22879157 | 0.003967039 | 0.003967039 |
| 1425092_at | Cdh10 | ENSMUSG00000022321 | 320873 | 3.281944768 | 3.038178309 | 0.243766459 | 1.18407992 | 0.004043584 | 0.004043584 |
| 1418649_at | Egln3 | \|ENSMUSG00000035105 | 112407 | 8.558848309 | 7.913266323 | 0.645581986 | 1.56437023 | 0.004519995 | 0.004519995 |
| 1460366_at | Eml3 | ENSMUSG00000071647 | 225898 | 7.542051934 | 6.985281848 | 0.556770086 | 1.47097231 | 0.004758655 | 0.004758655 |
| 1423626_at | Dst | ENSMUSG00000026131 | 13518 | 3.913358178 | 4.300789513 | -0.387431335 | -1.30806237 | 0.004931354 | 0.004931354 |
| 1427323_s_at | Wipi1 | ENSMUSG00000041895 | 52639 | 5.672041429 | 5.921793133 | -0.249751704 | -1.18900246 | 0.004931692 | 0.004931692 |
| 1427794_at | AJ242955 | | 58363 | 3.106623887 | 2.824001526 | 0.282622362 | 1.21640391 | 0.004961076 | 0.004961076 |
| 1422344_s_at | Tnfrsf10b | ENSMUSG00000022074 | 21933 | 4.383312428 | 4.119481269 | 0.263831159 | 1.2006629 | 0.005424567 | 0.005424567 |
| 1415832_at | Agtr2 | ENSMUSG00000068122 | 11609 | 2.5151512 | 2.302923442 | 0.212227757 | 1.15847568 | 0.005552982 | 0.005552982 |
| 1420735_at | Gabrr2 | \|ENSMUSG00000023267 | 14409 | 3.649305434 | 3.44163256 | 0.207672874 | 1.1548239 | 0.005776048 | 0.005776048 |
| 1418352_at | Hsd17b2 | ENSMUSG00000031844 | 15486 | 3.530858763 | 3.232555005 | 0.298303758 | 1.22969775 | 0.006138543 | 0.006138543 |
| 1419173_at | Acy1 | ENSMUSG00000023262 | 109652 | 7.350027228 | 6.815307292 | 0.534719935 | 1.4486609 | 0.006154995 | 0.006154995 |
| 1417047_at | Prom2 | ENSMUSG00000027376 | 192212 | 3.37313633 | 3.071736026 | 0.301400305 | 1.23233996 | 0.006445766 | 0.006445766 |
| 1434100_x_at | | \|ENSMUSG00000029167 | | 2.401771307 | 2.149576689 | 0.252194618 | 1.19101751 | 0.006453775 | 0.006453775 |
| 1417812_a_at | Lamb3 | ENSMUSG00000026639 | 16780 | 3.470801319 | 3.228651273 | 0.242150046 | 1.18275401 | 0.006477274 | 0.006477274 |

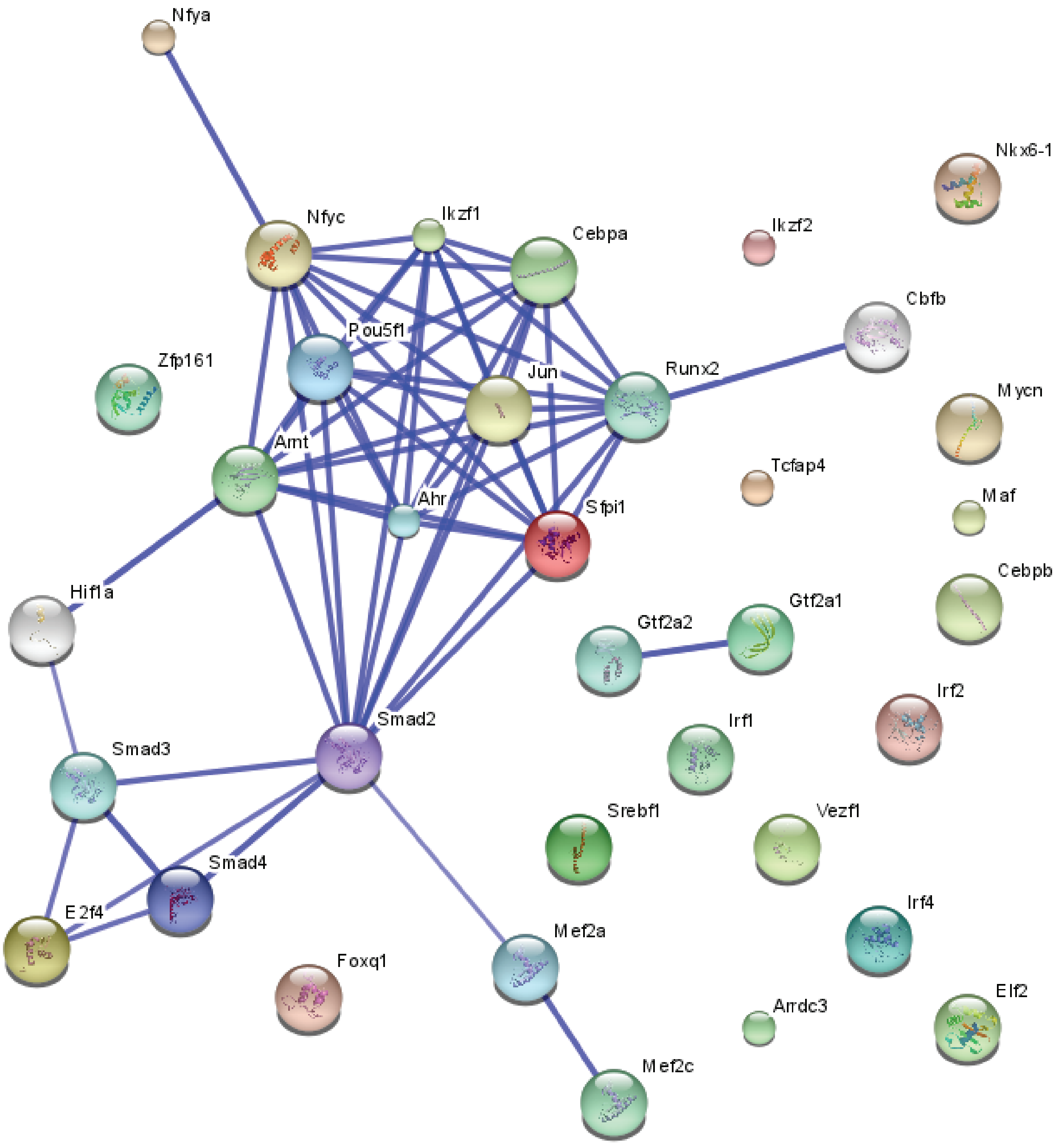

**Figure S1: Protein-protein interaction network of TFs** with TFBS enriched in genes with up-regulated expression in Nup98 translocations.

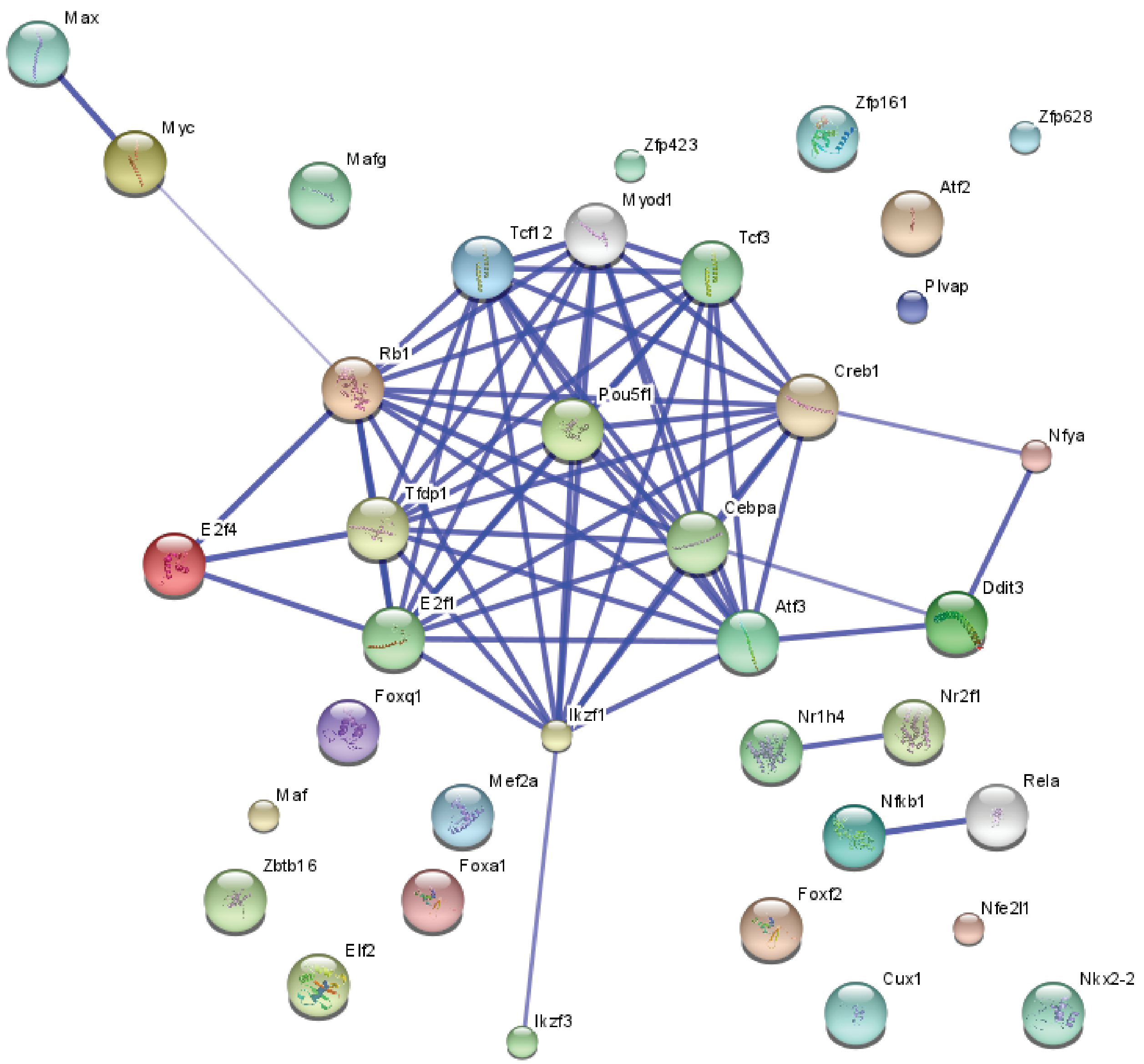

**Figure S1: Protein-protein interaction network of TFs** with TFBS enriched in genes with up-regulated expression in Nup98 translocations.